\documentclass[aps,prc,amsmath,superscriptaddress]{revtex4} 
\renewcommand{\baselinestretch}{1.35}
\begin{document} 

\title{\bf The scientific heritage of Richard Henry Dalitz, FRS 
(1925-2006)} 

\author{Ian J.R.~Aitchison} 
\email{I.Aitchison1@physics.ox.ac.uk} 
\affiliation{Rudolf Peierls Centre for Theoretical Physics, University of 
Oxford, Oxford OX1 3NP, United Kingdom} 
\affiliation{Stanford Linear Accelerator Center, Stanford University, 
Stanford, CA 94309, USA}

\author{Frank E.~Close}
\email{F.Close@physics.ox.ac.uk} 
\affiliation{Rudolf Peierls Centre for Theoretical Physics, University of 
Oxford, Oxford OX1 3NP, United Kingdom} 

\author{Avraham~Gal}
\email{avragal@vms.huji.ac.il} 
\affiliation{Racah Institute of Physics, The Hebrew University,
Jerusalem 91904, Israel}

\author{D. John~Millener} 
\email{millener@bnl.gov} 
\affiliation{Physics Department, Brookhaven National Laboratory, Upton, 
NY 11973, USA} 

\begin{abstract}

Professor Richard H. Dalitz passed away on January 13, 2006. He was almost 
81 years old and his outstanding contributions are intimately connected to 
some of the major breakthroughs of the 20th century in particle and nuclear 
physics. These outstanding contributions go beyond the Dalitz Plot, 
Dalitz Pair and CDD poles that bear his name. He pioneered the theoretical 
study of strange baryon resonances, of baryon spectroscopy in the quark model, 
and of hypernuclei, to all of which he made lasting contributions. His 
formulation of the ``$\theta-\tau$ puzzle" led to the discovery that parity 
is not a symmetry of the weak interactions. A brief scientific evaluation of 
Dalitz's major contributions to particle and nuclear physics is hereby 
presented, followed by a comprehensive list of his scientific publications, 
as assembled from several sources. The list is divided into two categories: 
the first, main part comprises Dalitz's research papers and reviews, 
including topics in the history of particle physics, biographies and 
reminiscences; the second part lists book reviews, public lectures and 
obituaries authored by Dalitz, and books edited by him. This provides 
the first necessary step towards a more systematic research of the Dalitz 
heritage in modern physics. 

The present 2016 edition updates the original 2006 edition, published in 
Nucl. Phys. A 771 (2006) 2-7, doi:10.1016/j.nuclphysa.2006.03.007, and 
8-25, doi:10.1016/j.nuclphysa.2006.03.008, by including for the first time 
a dozen or so of publications, found recently in a list submitted to the 
Royal Society by Dalitz in 2004, that escaped our attention in the original 
version. 

\end{abstract} 
\date{\today}
\maketitle 

\begin{acknowledgments} 

Special thanks for illuminating little known aspects of Dalitz's scientific 
career, and for assistance in identifying and locating references, go to 
Michael Berry, Roelof Bijker, Gerry Brown, Hugh Burkhardt, Don Davis, Andrzej 
Deloff, Norman Dombey, Evgenii Drukarev, Gary Goldstein, Ron Horgan, Dave 
Jackson, Harry Lipkin, Chris Llewellyn Smith, Jim Lowe, Ji\v{r}\'{\i} 
Mare\v{s}, Alan Martin, Gordon Moorhouse, Peter Norton, Hugh Osborn, Ken 
Peach, Willi Plessas, Michal Praszalowicz, Guruswamy Rajasekaran, Marc Rayet, 
Jonathan Rosner, Mikko Sainio, Madeleine Soyeur, Gerry Stone and Igor 
Strakovsky. IJRA is grateful for the assistance provided by the staff of the 
SLAC library. DJM acknowledges the special help extended by Kathryn Lancaster 
and Madeline Windsor of the BNL Research Library. He is supported by the US 
Department of Energy under Contract DE-AC02-98-CH10886. FEC and AG thank Nanci 
Carrothers of the Enrico Fermi Institute, University of Chicago, for providing 
relevant bibliographical material, and Rob Edgecock of the Rutherford 
Laboratory, for circulating an early version of the List. 

\end{acknowledgments}

\section{Dalitz's impact on 20th century physics} 
\noindent 
{\it by Frank E. Close and Avraham Gal} 
\newline
\newline
Dick Dalitz was born in Dimboola, 
Victoria Australia, on February 28th 1925, and gained B.A. and B.Sc. degrees 
in Mathematics and Physics in 1944 and 1945, respectively, from the 
University of Melbourne. He moved to Britain in 1946 to do his Ph.D. at 
Cambridge, and then worked at the University of Bristol before joining in 1949 
Rudolf Peierls in Birmingham where he subsequently became a Lecturer. He spent 
two years in the U.S. from 1953, holding research positions at Cornell and 
Stanford, visiting also Princeton and Brookhaven National Laboratory, 
and returned as a Reader in Mathematical Physics to the University of 
Birmingham for a year before becoming Professor of Physics in the Enrico Fermi 
Institute for Nuclear Studies and the Department of Physics at the University 
of Chicago. He moved to Oxford in 1963 as a Royal Society 
Research Fellow, the post he held until his retirement in 1990. 
Dick Dalitz made many outstanding contributions to particle physics, 
beyond the Dalitz Plot, Dalitz Pair and CDD (Castillejo, Dalitz, Dyson) poles 
that bear his name. He pioneered the theoretical study of strange baryon 
resonances, of baryon spectroscopy in the quark model, and of hypernuclei, 
to all of which he made lasting contributions. His formulation of the 
``$\theta-\tau$ puzzle" led to the discovery that parity is not a symmetry 
of the weak interactions. 
\newline
\newline
His Cambridge Ph.D. thesis, supervised by Nicholas Kemmer and completed in 
1950, was on ``Zero-zero transitions in nuclei". It evolved from the 
theoretical advice that Sam Devons sought on his novel nuclear experiments 
at Cambridge. Primarily it was a study of the well-known deexcitation by 
electron-positron pair emission of the $0^+$ first excited level at 6.05 MeV 
to the $0^+$ ground state of $^{16}$O. Emission of a real photon is forbidden 
due to angular momentum conservation, but it is allowed for a longitudinally 
polarised virtual photon that converts into an electron-positron pair. 
His work on the higher order radiative corrections in such pair emission 
transitions in nuclei rapidly bore fruit in particle physics in 1951 with 
his seminal contribution of ``Dalitz pairs" \cite{Dal51}.  
This is a form of internal conversion in $\pi^0 \to 2\gamma$ decay where one 
of the photons converts into an $e^+e^-$ pair, with a branching ratio 
$1.2 \times 10^{-2}$, or for both photons to convert into Dalitz pairs with 
a branching ratio of $3.1 \times 10^{-5}$. These Dalitz pairs were observed 
in photographic emulsion experiments, as pairs coming from the origin of 
a $\pi$ or $K$ interaction or decay and, to mention just two examples, 
were used to measure the parity of the $\pi^0$ \cite{Pla59} and to establish, 
by the decay of the $\Sigma^0$, that the $\Sigma^0$ and $\Lambda$ have the 
same parity \cite{Cou63}. 
\newline
\newline
It was his work with the $\tau$ meson that revolutionised particle physics. 
During his Ph.D. thesis he had spent a year working alongside Cecil Powell's 
cosmic ray group at Bristol and it was during this period that he took 
particular interest in the strange particles that were beginning to appear in 
cosmic rays and at particle accelerators. These included the first 
hyperfragment in 1953, which inspired a lifelong interest in hypernuclei, 
and the observation of two kinds of mesons named $\theta$ and $\tau$, 
with essentially the same masses and lifetimes. The positive-charge species, 
both now understood to be the same $K^+$ meson, were clearly distinguished by 
their decays $\theta^+ \to \pi^+\pi^0$ into two pions and 
$\tau^+ \to \pi^+\pi^+\pi^-$ into three pions. Conservation of parity in 
$\theta \to 2\pi$ would imply that it had $J^P = 0^+,1^-,2^+$..., and the 
question was whether the $\tau$ which decays to three pions could have any of 
these quantum numbers. In 1953, Dalitz looked at the decays of the $\tau$ into 
three pions and in doing so introduced what he would modestly call a phase 
space plot, but which is known throughout physics as the Dalitz plot 
\cite{Dal53,Dal54}. The distribution of events in this kinematical 
two-dimensional plot led to 
the conclusion that the $\tau$ had even spin and odd parity, such as $0^-$. 
Thus was born the $\theta - \tau$ puzzle: how could two mesons have the same 
masses and lifetimes and yet have different quantum numbers? The puzzle 
persisted for two years: Dalitz even musing to colleagues that perhaps the 
law of parity had to be abandoned, although all the evidence at the time 
appeared to say otherwise. It was T.D. Lee and C.N. Yang, who in 1956 
realised that the assumption of conserved parity in weak interactions, 
such as $\beta$ decay of nuclei, had not been tested, and it was the weak 
force that was at work in the $\theta - \tau$ decays. They were proved to be 
right, for which they won in 1957 a well-deserved Nobel Prize. 
Thus, the $\theta$ and the $\tau$ were the same $K$ meson. 

In his 1957 review {\it $K$ Mesons and Hyperons} \cite{Dal57}, Dalitz showed 
that their parity could be neatly derived from a strong-interaction 
production process; this established in due course that the $K$-meson parity 
is opposite to that of the $\Lambda$, precisely the same relation as between 
the parities of the $\pi$ meson and the nucleon. 
\newline 
\newline 
Dalitz's interests and contributions were not limited to electromagnetic 
and weak interactions. He made significant contributions to the strong 
interactions of the strange particles and their resonant states, soon 
reviewed in Reviews of Modern Physics \cite{Dal61} and in Annual Review of 
Nuclear Science \cite{Dal63}. In fact, as early as 1959 Dalitz and Tuan, 
by analysing the data on the strong interactions of $K^-$ mesons with protons, 
predicted the existence of an $I=0,~J^P=(1/2)^-$ `strange' resonance about 
20~MeV below the $K^-p$ threshold \cite{Dal59}. This $\Lambda(1405)$ 
resonance was discovered two years later, during Dalitz's sabbatical year in 
Berkeley, by the Alvarez team in the Berkeley hydrogen bubble chamber, 
studying the reaction $K^-p \to \Sigma + 3\pi$ for several charge states 
\cite{Als61}. The discovery of $\Lambda(1405)$ had been preceded by that of 
the $I=1,~J^P=(3/2)^+$ strange resonance $\Sigma(1385)$ \cite{Als60} in 
$K^-p \to \Lambda \pi^+\pi^-$ interactions, displaying the three final 
state particles on a Dalitz plot. The $p$-wave $\pi \Lambda$ resonance 
$\Sigma(1385)$ provided a straightforward generalisation of the non-strange 
$p$-wave $\pi N$ $\Delta(1238)$ resonance discovered by Fermi and coworkers 
in 1952 into the strange sector, and within a few years led to a complete 
SU(3) decuplet; by contrast, the $\Lambda(1405)$ required a different 
dynamical origin. The proximity of this $s$-wave $\pi \Sigma$ resonance 
to the $\bar K N$ threshold suggested that it can be generated by the 
$\bar K N-\pi\Sigma$ inter-hadron forces, and this was shown in 1967 by 
Dalitz, Wong and Rajasekaran to be possible within a dynamical model of 
SU(3)-octet vector-meson exchange \cite{DWR67}. 
The underlying vector mesons $\rho, \omega, K^{\star}, \phi$, which were 
discovered in the years 1960-61, relying heavily on Dalitz plots for some of 
these, were unknown when Dalitz and Tuan predicted the $\Lambda(1405)$. 
In the years to follow, Dalitz repeatedly considered the completeness 
of this dynamical picture, whether or not the $\Lambda(1405)$ $S$-matrix 
pole due to the inter-hadron forces need not be augmented by a CDD pole 
arising from inter-quark forces upon allowing for an intermediate $uds$ 
configuration. It is here that the earlier CDD discussion \cite{CDD56} 
found a fertile physical ground. 
\newline 
\newline  
The discoveries of these, and other baryon and meson resonances as well, 
paved the way for Gell Mann and Ne'eman in 1961 to introduce SU(3) (flavour) 
symmetry into particle physics and the idea that a more fundamental level of 
reality existed in what Gell Mann called quarks in 1964. It was initially 
unclear whether these fractionally-charged quarks were just a mathematical 
convenience or were themselves real particles. It was around this time that 
Dalitz returned to Britain in 1963 joining Rudolf Peierls at Oxford as Royal 
Society research professor. 

Dalitz took the idea of physical quarks very 
seriously and proposed that they were the basic blocks of baryons and mesons. 
In his remarkable Les Houches summer school notes (1965), Tokyo summer 
lectures (1966) and Berkeley conference talk (1966), he showed how the idea 
explained properties of the proton and neutron, such as their response to 
a magnetic field and then took his more radical step of exciting the quarks 
into different energy states for mesons and baryons, following the established 
rules of non-relativistic quantum mechanics. To explain the pattern of the 
baryon SU(3) octet and decuplet representations, embedded in the $\bf 56$ 
dimensional SU(6) (spin-flavour) {\it symmetric} representation, required 
assuming what became known as a symmetric quark model, in defiance of the 
established antisymmetry for fermions. While this was developed by several 
others, it was Dalitz who analysed the three body state with internal orbital 
angular momentum excitations, using his earlier experience with the 
$\tau \to 3 \pi$ decay, leading to a $\bf 70$ dimensional SU(6) representation 
with negative parity as the first excited states, as seen empirically, and 
with higher excitations in $\bf 56$, $\bf 70$ and $\bf 20$ dimensional 
representations of SU(6) with negative or positive parity. Over the following 
decades many other resonances were discovered, for both baryons and mesons, 
in many cases by application of Dalitz plots, such that the non-relativistic 
quark model, with QCD motivated SU(3) (colour) effects incorporated later, 
became established as a description of what had hitherto been a menagerie 
of particles. 

While the quark model description of the low lying baryons has 
proved generally successful, its classification for the $\Lambda(1405)$ as 
a $uds$ SU(3)-singlet $P_{\frac{1}{2}}$ poses a serious problem. It was shown 
by Close and Dalitz \cite{CDa81} that 
a consistent application of the Fermi-Breit interaction in QCD produces 
a spin-orbit splitting that places this $P_{\frac{1}{2}}$ state higher than 
its $LS$ doublet partner $P_{\frac{3}{2}}$, hence above the $D$-wave 
$\bar K N$ resonance $\Lambda(1520)$ which is fitted well by this assignment. 
Even if one accepts that the structure of $\Lambda(1405)$ is dominated by 
inter-hadron forces, as in the original scenario put forward by Dalitz and 
Tuan, there remains a lingering doubt about a missing candidate for the quark 
model assignment. 
\newline 
\newline 
Dalitz pioneered the theoretical study of hypernuclei, atomic nuclei in which 
a $\Lambda$ hyperon is bound and which are as long lived as the $\Lambda$'s 
free-space lifetime of order $10^{-10}$~sec. $\Lambda$ hypernuclei, produced 
by $K^-$ mesons stopping in emulsion and identified by their decay products, 
were discovered by Danysz and Pniewski in 1953, and have been produced 
systematically in accelerators for the last 30 years using $K^-$ and $\pi^+$ 
beams to convert a bound neutron into a bound $\Lambda$. 
[Danysz and Pniewski also headed the discovery in 1963 of the first of several 
known to date double-$\Lambda$ hypernuclei, from which Dalitz immediately 
provided the first estimate on the $\Lambda-\Lambda$ interaction; he would 
come back in 1989 to this issue in connection with the then argued stability 
of a dihyperon $H$ made out of $uuddss$ quarks in a different configuration 
than just two $\Lambda$s (each one made out of $uds$)]. Hypernuclei provide 
unique information on the strangeness degree of freedom for studies of dense  
hadronic or quark matter in stars. 

Dalitz's first published work on 
hypernuclei dates back to 1955, dealing with charge independence in light 
hypernuclei. In a series of works covering three decades, he used the main 
$\Lambda \to p \pi^-$ weak-decay mode of the light species ($A \lesssim 13$) 
studied in emulsion and bubble chambers to determine their ground-state spins 
and, thereby to gain information on the spin dependence of the $\Lambda N$ 
force. But his early outstanding contribution to weak interactions in 
hypernuclei, together with Martin Block \cite{BDa63}, was to formulate the 
phenomenology of non-mesonic $\Lambda N \to NN$ weak-interaction decay modes 
that dominate the decays of heavier hypernuclei, 
a process that cannot be studied on free baryons and which offers new systems, 
$\Lambda$ hypernuclei, for exploring the little understood $\Delta I = 1/2$ 
rule in non-leptonic weak interactions. 

Another important contribution, in the 1970s, following the introduction of 
shell-model techniques by Gal, Soper and Dalitz, was to chart together with 
Avraham Gal \cite{DGa78} the production and electromagnetic decay schemes 
anticipated for excited states in light hypernuclei in order to derive the 
complete spin dependence of the $\Lambda N$ interaction effective in these 
hypernuclei. This work, developed further together with John Millener and 
Carl Dover \cite{MGD85}, served as a useful guide to the hypernuclear 
$\gamma$-ray measurements completed in the last few years, at KEK and BNL 
\cite{HTa06}, which yielded full determination of the spin dependence in the 
low-lying spectrum. 

Dalitz and Gal also studied the extension of Wigner's 
spin-isospin SU(4) symmetry in light nuclei to SU(6) approximate symmetry in 
light hypernuclei \cite{DGa76}, following a suggestion by Kerman and Lipkin 
to generalise isospin SU(2) into Sakata SU(3) symmetry for 
the $p,n,\Lambda$ constituents of hypernuclei. This resulted 
in a unique feature named by them `supersymmetry' (before this attribute 
was taken up in particle physics), holding for particularly symmetric 
$\Lambda$-hypernuclear excited states in which the $\Lambda$ occupies 
a quantum state forbidden for nucleons owing to the Pauli principle. 
Such states have been observed in $_{\Lambda}^9{\rm Be}$ at KEK \cite{HTa06} 
and bear important information on the effective $\Lambda N$ interaction in 
the $p$ shell. 

In later years Dalitz, mostly with Don Davis and Toshio Motoba, 
focused attention to unbound $\Lambda$-hypernuclear levels 
that had been precisely studied in nuclear emulsions by following their 
proton emission tracks. With Andrzej Deloff he collaborated 
on $K^-$ interactions in deuterium and helium. 
He was the recognised leader of this field of research for 50 years, 
and it is instructive to read through his own recollection of events in his 
last published paper \cite{Dal05} which is based on a talk 
given in HYP2003 at Jlab. 
\newline 
\newline 
He was also intimately involved with the identification of the top quark 
where he thought about the problems of how one might identify it from the 
decay processes that seemed most natural for it. With Gary Goldstein he worked 
out a geometrical method by which experimental data could be used to deduce 
the top mass. They applied the method to an early possible event from Fermilab 
and concluded that if this event indeed signalled top production, the top 
quark mass most probably exceeded 130 GeV \cite{DGo92}. 
This was regarded as an unexpectedly large value at the time. 
This one event might not even have been due to a top quark and the 
confirmation could only be decided on the basis of a large number of observed 
events, all of them being consistent with a unique mass. 
This was the case later when two experimental groups came to conclude 
that the top quark mass was about 180 GeV. 
\newline 
\newline  
Dalitz worked close to data and was greatly admired by experimentalists. 
He brought scholars to Oxford, which became a centre for the quark model, 
and trained generations of students, including Chris Llewellyn Smith a future 
Director-General of CERN. Following retirement he remained an inspirational 
figure to students new and old, continuing to work on theoretical physics 
with undiminished enthusiasm. With his death, international physics has 
lost a major figure and Britain one of its greatest unsung scientists. 
As the next phase in the quest for the ultimate nature of reality begins 
at CERN's Large Hadron Collider, it is likely that evidence for Higgs Bosons, 
supersymmetric particles, or whatever other surprises may await, 
may be revealed by Dalitz plots. 
 
\renewcommand{\baselinestretch}{1.00}

\newpage

\section{R.H. Dalitz -- main list of publications} 
\noindent 
{\it by I.J.R. Aitchison, F.E. Close, A. Gal and D.J. Millener} 
\begin{enumerate}

\item R.H. Dalitz, Some mathematical aspects of compressible flow, Australian 
Council for Aeronautics Report (ACA-20, Melbourne, January 1946) 39 pp. 
\item R.H. Dalitz, On higher Born approximations in potential scattering, 
Proc. Roy. Soc. A 206 (1951) 509-520. 
\item R.H. Dalitz, On radiative corrections to the angular correlation in 
internal pair creation, Proc. Roy. Soc. A 206 (1951) 521-538. 
\item R.H. Dalitz, On an alternative decay process for the neutral 
$\pi$-meson, Letters to the Editor, Proc. Phys. Soc. A 64 (1951) 667-669. 
\item R.H. Dalitz, D.G. Ravenhall, On the Tomanaga method for intermediate 
coupling in meson field theory, Phil.~Mag. 42 (1951) 1378-1383. 
\item R.H. Dalitz, On polarized particle beams, Proc. Phys. Soc. A 65 (1952) 
175-178. 
\item R.H. Dalitz, Some features of the deuteron stripping process, 
Proc. Phys. Soc. A 66 (1953) 28-32. 
\item R.H. Dalitz, The decay of the $\tau$-meson, Proc. Phys. Soc. A 66 
(1953) 710-713. 
\item R.H. Dalitz, The modes of decay of the tau-meson, in {\it Proc. Third 
Int'l Conf. Cosmic Rays, Bagn\'{e}res-de-Bigorre, France 1953}, pp. 236-239. 
\item R.H. Dalitz, On the analysis of $\tau$-meson data and the nature of 
the $\tau$-meson, Phil. Mag. 44 (1953) 1068-1080. 
\item R.H. Dalitz, Decay of $\tau$ mesons of known charge, Phys. Rev. 94 
(1954) 1046-1051. 
\item R.H. Dalitz, The Sachs exchange moment, Phys. Rev. 95 (1954) 799-800. 
\item R.H. Dalitz, The analysis of $\tau$-meson decay, in {\it Proc. 4th 
Annual Rochester Conf. High Energy Nuclear Physics, Rochester 1954}, 
eds. H.P. Noyes, E.M. Hafner, J. Klarmann, A.E. Woodruff (Interscience, 
New York, 1954) pp. 83-84. 
\item R.H. Dalitz, F.J. Dyson, Renormalization of the new Tamm-Dancoff 
theory of meson-nucleon scattering, Phys.~Rev. 99 (1955) 301-314. 
\item R.H. Dalitz, Radiative $\tau$-meson decay, Phys. Rev. 99 (1955) 915-920. 
\item R.H. Dalitz, Charge independence in light hyperfragments, 
Phys. Rev. 99 (1955) 1475-1477. 
\item M.H. Kalos, R.H. Dalitz, Pion-nucleon scattering calculations in the 
Tamm-Dancoff theory, Phys. Rev. 100 (1955) 1515-1522. 
\item R.H. Dalitz, Meson phenomena and the meson theory, Prog. Nucl. Phys. 
4 (1955) 95-141. 
\item R.H. Dalitz, Analysis of $\tau$ meson decays, in {\it Proc. 5th Annual 
Rochester Conf. High Energy Nuclear Physics, Rochester 1955}, eds. H.P. Noyes, 
E.M. Hafner, G. Yekutieli, B.J. Raz (Interscience, New York, 1955) pp. 140-142. 
\item L. Castillejo, R.H. Dalitz, F.J. Dyson, Low's scattering equation 
for the charged and neutral scalar theories, Phys. Rev. 101 (1956) 453-458. 
\item R.H. Dalitz, Isotopic spin changes in $\tau$ and $\theta$ decay, 
Proc. Phys. Soc. A 69 (1956) 527-540. 
\item R.H. Dalitz, M.K. Sundaresen, H.A. Bethe, A singular integral equation
in the theory of meson nucleon scattering, Proc. Cambridge Phil. Soc. 52 
(1956) 251-272. 
\item R.H. Dalitz, Nature of the $\Lambda^0$-nucleon force from binding 
energies of light hyperfragments, in {\it Proc. 6th Annual Rochester Conf. 
High Energy Nuclear Physics, Rochester 1956}, eds. J. Ballam, V.L. Fitch, 
T. Fulton, K. Huang, R.R. Rau, S.B. Treiman (Interscience, New York, 1956) 
pp. V 40-43. 
\item R.H. Dalitz, Present status of $\tau$ spin-parity, in {\it Proc. 6th 
Annual Rochester Conf. High Energy Nuclear Physics, Rochester 1956}, eds. 
J. Ballam, V.L. Fitch, T. Fulton, K. Huang, R.R. Rau, S.B. Treiman 
(Interscience, New York, 1956) pp. VIII 19-23. 
\item R.H. Dalitz, D.R. Yennie, Pion production in electron-proton collisions, 
Phys. Rev. 105 (1957) 1598-1615. 
\item R.H. Dalitz, $K$-mesons and hyperons: their strong and weak 
interactions, Rep. Prog. Phys. 20 (1957) 163-303. 
\item R.H. Dalitz, Eight lectures concerning the strong interactions of 
strange particles, given at Brookhaven National Laboratory, July-August 1957 
(BNL-3405, Upton, 1957) 72 pp.  
\item R.H. Dalitz, Theoretical interpretation of $\pi$ production, in {\it 
Proc. 7th Annual Rochester Conf. High Energy Nuclear Physics, Rochester 1957}, 
eds. G. Ascoli, G. Feldman, L.J. Koester, R. Newton, W. Riesenfeld, M. Ross, 
R.G. Sachs (Interscience, New York, 1957) pp. II 61-64. 
\item R.H. Dalitz, Theoretical analysis of hyperfragments, in {\it Proc. 7th 
Annual Rochester Conf. High Energy Nuclear Physics, Rochester 1957}, 
eds. G. Ascoli, G. Feldman, L.J. Koester, R. Newton, W. Riesenfeld, M. Ross, 
R.G. Sachs (Interscience, New York, 1957) pp. VIII 12-14.
\item R.H. Dalitz, B.W. Downs, Remarks on the hypertriton, Phys. Rev. 110 
(1958) 958-965. 
\item R.H. Dalitz, B.W. Downs, Hypernuclear binding energies and the 
$\Lambda$-nucleon interaction, Phys. Rev. 111 (1958) 967-986. 
\item R.H. Dalitz, Parity nonconservation in the decay of free and bound 
$\Lambda$ particles, Phys. Rev. 112 (1958) 605-613. 
\item R.H. Dalitz, Strange particle interactions - theoretical, in 
{\it Proc. 1958 Annual Int'l Conf. High Energy Physics, CERN}, ed. B. Ferreti 
(CERN, Geneva, 1958) pp. 187-203. 
\item B.W. Downs, R.H. Dalitz, Analysis of the $\Lambda$-hypernuclear 
three-body systems, Phys. Rev. 114 (1959) 593-602. 
\item R.H. Dalitz, S.F. Tuan, A possible resonant state in pion-hyperon 
scattering, Phys. Rev. Lett. 2 (1959) 425-428. 
\item R.H. Dalitz, Strange particle decay processes and the Fermi interaction, 
Rev. Mod. Phys. 31 (1959) 823-833. 
\item R.H. Dalitz, S.F. Tuan, The energy dependence of low energy $K^-$-proton 
processes, Ann. Phys. 8 (1959) 100-118. 
\item R.H. Dalitz, L. Liu, Pionic decay modes of light 
$\Lambda$ hypernuclei, Phys. Rev. 116 (1959) 1312-1321. 
\item R.H. Dalitz, The decay of strange particles, in {\it Weak Interactions, 
Proc. Int'l School of Physics ``Enrico Fermi", Course XI, 1959}, 
ed. L.A. Radicati (Nicola Zanichelli, Bologna, 1960) pp. 299-335. 
\item R.H. Dalitz, Hyperfragments and hyperon-nucleon interaction, 
in {\it Proc. Ninth Int'l Annual Conf. High Energy Physics, Kiev 1959} 
(Academy of Sciences USSR, Moscow, 1960) Vol. 1, pp. 587-606. 
\item R.H. Dalitz, Elementary particle physics today, Nucleus (Sydney) 6(2) 
(1960) 15-20.
\item R.H. Dalitz, S.F. Tuan, The phenomenological description of 
$\bar K$-nucleon reaction processes, Ann. Phys. 10 (1960) 307-351. 
\item R.H. Dalitz, $\bar K$-nucleon bound-state interpretation of the 
1385-Mev $\pi$-$\Lambda$ resonance, Phys. Rev. Lett. 6 (1961) 239-241. 
\item R.H. Dalitz, D.H. Miller, Bose statistics and $Y^{\star}$ production 
and decay in $K^-$-$p$ collisions, Phys. Rev. Lett. 6 (1961) 562-567. 
\item R.H. Dalitz, On the strong interactions of the strange particles, 
Rev. Mod. Phys. 33 (1961) 471-492. 
\item R.H. Dalitz, Hypernuclei and hyperon-nucleon interaction, in 
{\it Proc. Rutherford Jubilee Int'l Conf., Manchester 1961}, ed. J.B. Birks 
(Heywood, London, 1961) pp. 103-137.
\item R.H. Dalitz, Three lectures on elementary particle resonances, 
given at Brookhaven National Laboratory, 13-15 December, 1961 
(BNL-735 (T-264), Upton, 1962) 96 pp. 
\item R.H. Dalitz, Some topics in strange particle physics, 
in {\it The Aix-en-Provence Int'l Conf. Elementary Particles, 1961}, 
eds. E. Cremieu-Alcan,  P. Falk-Vairant, O. Lebey (C.E.N., Saclay, 1962) 
Vol. 2, pp. 151-175. 
\item R.H. Dalitz, Strange Particles and Strong Interactions 
(Oxford University Press, London, 1962) 187 pp. 
\item R.H. Dalitz, G. Rajasekharan, The spins and lifetimes of the light 
hypernuclei, Phys. Lett. 1 (1962) 58-60. 
\item R. Chand, R.H. Dalitz, Charge-independence in $K^-$-deuterium capture 
reactions, Ann. Phys. 20 (1962) 1-19. 
\item R.H. Dalitz, Lectures on the properties and the symmetry theories of 
the resonant states, delivered at the {\it Summer School in Theoretical 
Physics, 1962, Bangalore}, notes taken by S.N. Biswas, V. Gupta (Tata Inst. 
Fundamental Research, Bombay, 1962) 112 pp. 
\item R.H. Dalitz, The pion-hyperon couplings, the $\overline K N$ 
interactions, and the $Y^{\star}$ resonant states, in {\it Proc. Int'l Conf. 
High-Energy Physics at CERN}, ed. J. Prentki (CERN, Geneva, 1962) pp. 391-401. 
\item R.H. Dalitz, Strange-particle resonant states, Annu. Rev. Nucl. Sci. 13 
(1963) 339-430. 
\item R.H. Dalitz, R. Levi Setti, Some possibilities for unusual light 
hypernuclei, Nuovo Cimento 30 (1963) 489-501. 
\item M.M. Block, R.H. Dalitz, Structure of the weak interaction 
$\Lambda + N \to N + N$, Phys. Rev. Lett. 11 (1963) 96-100. 
\item R.H. Dalitz, The hypernucleus $_{\Lambda}^8{\rm Li}$ and the decay 
process $_{\Lambda}^8{\rm Li} \to \pi^- + ^4{\rm He} + ^4{\rm He}$, 
Nucl. Phys. 41 (1963) 78-91. 
\item R.H. Dalitz, The $\Lambda\Lambda$-hypernucleus and the 
$\Lambda-\Lambda$ interaction, Phys. Lett. 5 (1963) 53-56. 
\item R.H. Dalitz, G. Rajasekaran, Resonance poles and mass differences 
within unitary multiplets, Phys. Lett. 7 (1963) 373-377. 
\item R.H. Dalitz, The systematics of baryon and meson states, in 
{\it Proc. Athens Topical Conf. on Recently Discovered Resonant Particles}, 
ed. B.A. Munir (Ohio University, Athens, 1963) 32 pp. 
\item R.H. Dalitz, Particles, isobars and unitary symmetry, in {\it Proc. 
Sienna Int'l Conf. on Elementary Particles}, eds. G. Bernardini, G. Puppi 
(Italian Physical Society, Bologna, 1963) Vol. 2, pp. 171-187. 
\item R.H. Dalitz, The $\Delta I =1/2$ rule for non-leptonic 
strangeness-changing decay processes, in {\it Proc. Int'l Conf. Fundamental 
Aspects of Weak Interactions, Brookhaven National Laboratory, September 1963}, 
eds. G.C. Wick, W.J. Willis (BNL-837 (C-39), Upton, 1964) pp. 378-409.
\item R.H. Dalitz, The strong and weak interactions of bound $\Lambda$ 
particles, in {\it Proc. Int'l Conf. Hyperfragments, St.~Cergue 1963}, 
ed. W.O. Lock (CERN Report 64-1, Geneva, 1964) pp. 147-172. 
\item R.H. Dalitz, The outlook in hypernuclear physics, in {\it Proc. Int'l 
Conf. Hyperfragments, St.~Cergue 1963}, ed.~W.O. Lock 
(CERN Report 64-1, Geneva, 1964) pp. 201-214. 
\item R.H. Dalitz, G. Rajasekaran, The binding of 
$\Lambda\Lambda$-hypernuclei, Nucl. Phys. 50 (1964) 450-464. 
\item R.H. Dalitz, F. von Hippel, Electromagnetic $\Lambda-\Sigma^0$ mixing 
and charge symmetry for the $\Lambda$-hyperon, Phys. Lett. 10 (1964) 153-157. 
\item R.H. Dalitz, F. von Hippel, The charge-exchange contributions to $\pi^+$ 
emission in hypernuclear decay, Nuovo Cimento 34 (1964) 799-806. 
\item R.H. Dalitz, Properties of the weak interactions, in 
{\it Weak Interactions and High-Energy Neutrino Physics, Proc. Int'l School 
of Physics ``Enrico Fermi", Course XXXII, 1964}, ed. T.D. Lee (Academic Press, 
New York, 1966) pp. 206-270. 
\item R.H. Dalitz, The production and decay of resonant states, in 
{\it Strong Interactions, Proc. Int'l School of Physics ``Enrico Fermi", 
Course XXXIII, 1964}, ed. L.W. Alvarez (Academic Press, New York, 1966) 
pp. 141-188. 
\item R.H. Dalitz, Nuclear Interactions of the Hyperons (Oxford University 
Press, London, 1965) 106 pp. 
\item R.H. Dalitz, Weak interactions of the muon, Proc. Roy. Soc. A 285 (1965) 
229-247. 
\item R.H. Dalitz, Deviations from unitary symmetry for resonant states, 
Proc. Roy. Soc. A 288 (1965) 183-199.  
\item R.H. Dalitz, D.G. Sutherland, $X^0-\eta$ mixing and some radiative meson
decay processes, Nuovo Cimento 37 (1965) 1777-1784; Erratum, Nuovo Cimento 38 
(1965) 1945. 
\item R.H. Dalitz, T.W. Thacker, Electromagnetic radii of $^3{\rm H}$ and 
$^3{\rm He}$, Phys. Rev. Lett. 15 (1965) 204-207; Erratum, Phys. Rev. Lett. 15 
(1965) 334. 
\item R.H. Dalitz, R.G. Moorhouse, The $P_{11}$ pion-nucleon phase shift - 
a resonant state at 1400 MeV?, Phys. Lett. 14 (1965) 159-162; Erratum, 
Phys. Lett. 14 (1965) 356.  
\item R.H. Dalitz, Quark models for the `Elementary Particles', in 
{\it High Energy Physics, Les Houches 1965}, eds. C.~DeWitt, M. Jacob 
(Gordon and Breach, New York, 1965) pp. 251-323. 
\item R.H. Dalitz, Hypernuclear interactions, in {\it Proc. Topical Conf. on 
the use of Elementary Particles in Nuclear Research, Brussels 1965}, 
(IISN Univ. Libre de Bruxelles, Brussels, 1966) pp. 293-370. 
\item R.H. Dalitz, Resonant states and strong interactions, in {\it Proc. 
Oxford Int'l Conf. Elementary Particles, 1965}, eds. R.G. Moorhouse, 
A.E. Taylor, T.R. Walsh (Rutherford High Energy Laboratory, Chilton, 1966) 
pp. 157-181. 
\item M. Rayet, R.H. Dalitz, Lifetime of $_{\Lambda}^3{\rm H}$, Nuovo Cimento 
A 46 (1966) 786-794. 
\item R.H. Dalitz, Introductory remarks on study of pion-nucleon interaction, 
Proc. Roy. Soc. A 289 (1966) 442-448. 
\item R.H. Dalitz, D.G. Sutherland, M1 photo-excitation $N \to N^{\star}$ 
and SU(6) symmetry, Phys. Rev. 146 (1966) 1180-1185. 
\item R.H. Dalitz, Constraints on the statistical tensor for low-spin 
particles produced in strong interaction processes, Nucl. Phys. 87 (1966) 
89-99. 
\item R.H. Dalitz, Baryonic resonances, in {\it Proc. 1966 Midwest 
Conf. Theoretical Physics, Indiana University}, ed. D.B. Lichtenberg 
(Indiana University, Bloomington, 1966) pp. 94-117.
\item R.H. Dalitz, Symmetries and strong interactions, in {\it Proc. XIII 
Int'l Conf. High Energy Physics, Berkeley 1966}, ed. M. Alston-Garnjost 
(University of California Press, Berkeley and Los Angeles, 1967) pp.~215-236. 
\item R.H. Dalitz, Hypernuclear physics, in {\it Interaction of High-Energy 
Particles with Nuclei, Proc. Int'l School of Physics ``Enrico Fermi", 
Course XXXVIII, 1966}, ed. T.E.O. Ericson (Academic Press, New York, 1967) 
pp.~89-131. 
\item R.H. Dalitz, Quark substructure for mesonic and baryonic states, in 
{\it Proc. Liperi Summer School in Theoretical Physics, 1966}, 
ed. N. Mustelin, Vol. A: Elementary Particle Physics (Research Inst. Theor. 
Phys. University of Helsinki, Helsinki, 1967) pp. 170-263. 
\item R.H. Dalitz, Quark substructure for mesonic and baryonic states, in 
{\it 1966 Tokyo Summer Lectures in Theoretical Physics: Part II, Elementary 
Particle Physics}, eds. G. Takeda, A. Fujii (Syokabo, Tokyo; 
and W.A. Benjamin, New York, 1967) pp. 56-105. 
\item R.H. Dalitz, Some polarized target experiments for elementary physics, 
in {\it Proc. Int'l Conf. on Polarized Targets and Ion Sources, Saclay 1966} 
(C.E.N., Saclay, 1967) pp. 261-272.
\item R.H. Dalitz, A. Gal, Isomeric states in $_{\Lambda}^7{\rm He}$, 
Nucl. Phys. B 1 (1967) 1-6. 
\item R.H. Dalitz, T.C. Wong, G. Rajasekaran, Model calculation for 
$Y_0^{\star}$(1405) resonance state, Phys. Rev. 153 (1967) 1617-1623. 
\item R.H. Dalitz, Some problems with the quark model for elementary 
particles, in {\it Methods and Problems in Theoretical Physics, in Honour of 
R.E. Peierls, Birmingham 1967}, ed. J.E. Bowcock (North Holland, Amsterdam, 
1970) pp. 343-380. 
\item R.H. Dalitz, Excited baryons and the baryonic supermultiplets, 
in {\it Pion Nucleon Scattering, Proc. Irvine 1967}, eds. G.L. Shaw, 
D.Y. Wong (Wiley, New York, 1969) pp. 187-207; and in {\it Developments 
in the Quark Theory of Hadrons, Vol. 1: 1964-1978}, eds. D. B. Lichtenberg, 
S. P. Rosen (Hadronic Press, Nonantum USA, 1980) pp. 316-336.  
\item R.H. Dalitz, Hadron spectroscopy, in {\it Proc. II Hawaii Topical Conf. 
Particle Physics, 1967}, eds. S. Pakvasa, S.F. Tuan (University of Hawaii 
Press, Honolulu, 1968) pp. 325-466. 
\item L. Lovitch, S. Rosati, R. Dalitz, The stability of 
$_{\Lambda}^6{\rm Li}$, Nuovo Cimento A 53 (1968) 301-308; Erratum, Nuovo 
Cimento A 53 (1968) 1059.   
\item R.H. Dalitz, Mesonic resonance states, in {\it Meson Spectroscopy, 
Proc. Conf. Philadelphia 1968}, eds. C. Baltay, H.A. Rosenfeld (Benjamin, 
New York, 1968) pp. 497-584. 
\item R.H. Dalitz, Quarks, the hadronic sub-units? in {\it Contemporary 
Physics: Trieste Symposium 1968}, eds. L. Fonda, A. Salam (IAEA, Vienna, 1969) 
Vol. II, pp. 219-247. 
\item R.H. Dalitz, Some nuclear aspects of $\Lambda$-hypernuclear physics, 
in {\it Nuclear Physics, Proc. 1968 Les Houches Summer School of Theoretical 
Physics}, eds. C. DeWitt, V. Gillet (Gordon and Breach, New York, 1969) 
pp.~701-787. 
\item R.H. Dalitz, The present problems and future outlook in 
$\Lambda$-hypernuclear physics, in {\it Proc. Int'l Conf. Hypernuclear 
Physics, ANL 1969}, eds. A.R. Bodmer, L.G. Hyman (ANL, 1969) 708-747. 
\item R.H. Dalitz, Some comments on the quark model, concluding lecture, in 
{\it Symmetries and Quark Models, Proc. Int'l Conf. Wayne State Univ., 
Detroit 1969}, ed. R. Chand (Gordon and Breach, New York, 1970) pp. 355-401. 
\item R.H. Dalitz, Parity, charge conjugation, and time reversal, in {\it 
Proc. Int'l Mendeleev Congress, Turin 1969}, ed. M. Verde 
(Vincenzo Bona, 1971) pp. 341-375. 
\item R.H. Dalitz, Survey of the quark model for hadronic states, in {\it 
Proc. X Cracow School of Theoretical Physics, Zakopane 1970} (Report INP-713, 
Cracow, 1970) Vol. II, pp. 1-5. 
\item R.H. Dalitz, Introductory remarks concerning duality, Proc. Roy. Soc. 
A 318 (1970) 245-255. 
\item R.H. Dalitz, R.G. Moorhouse, What is resonance? Proc. Roy. Soc. A 318 
(1970) 279-298. 
\item A. Gal, J.M. Soper, R.H. Dalitz, Shell-model analysis of $\Lambda$ 
binding energies for p-shell hypernuclei. I. Basic formulas and matrix 
elements for $\Lambda N$ and $\Lambda NN$ forces, Ann. Phys. 63 (1971) 53-126. 
\item J.T. Londergan, R.H. Dalitz, Spin-orbit coupling for 
$\Lambda-^4$He scattering using a one-boson-exchange model for the 
$\Lambda N$ interaction, Phys. Rev. C 4 (1971) 747-764. 
\item R.H. Dalitz, Low energy hadron spectroscopy, in {\it Proc. Amsterdam 
Int'l Conf. Elementary Particles 1971}, eds. A.G. Tenner, M.J.G. Veltman 
(North-Holland, Amsterdam, 1972) pp. 201-264.
\item A. Gal, J.M. Soper, R.H. Dalitz, Shell-model analysis of $\Lambda$ 
binding-energies for p-shell hypernuclei. II. Numerical fitting, 
interpretation, and hypernuclear predictions, Ann. Phys. 72 (1972) 445-488. 
\item J.T. Londergan, R.H. Dalitz, Spin-orbit coupling for 
$\Lambda-^4$He scattering using a one-boson-exchange model for the 
$\Lambda N$ interaction. II., Phys. Rev. C 6 (1972) 76-86. 
\item R.H. Dalitz, R.C. Herndon, Y.C. Tang, Phenomenological study of s-shell
hypernuclei with $\Lambda N$ and $\Lambda NN$ potentials, Nucl. Phys. B 47 
(1972) 109-137. 
\item R.H. Dalitz, Relativistic systems and the Glauber theory of high energy 
scattering in the quark model, Summary of lectures, in {\it Proc. XII Cracow 
School of Theoretical Physics, Zakopane 1972} (Report INP-806, Cracow, 1972) 
pp. 83-86. 
\item R.H. Dalitz, Baryon spectroscopy with the quark model, in {\it 
Baryon Resonances - 73, Proc. Conf. Purdue University}, ed. E.C. Fowler 
(Purdue University, West Lafayette, 1973) pp. 393-416. 
\item R. Horgan, R.H. Dalitz, Baryon spectroscopy and quark shell-model. 
I. The framework, basic formulae, and matrix elements, Nucl. Phys. B 66 (1973) 
135-172; Erratum, Nucl. Phys. B 71 (1974) 546-547. 
\item R.H. Dalitz, Hypernuclear physics and its immediate problems, in 
{\it Proc. Summer Study Meeting on Nuclear and Hypernuclear Physics with 
Kaon Beams, BNL 1973}, ed. H. Palevsky (BNL Report 18335, Upton, 1973) 
pp. 1-63. 
\item R.H. Dalitz, The three-quark shell model and the spectrum of baryonic 
states, in {\it Hadron Interactions at Low Energies, Proc. Triangle Meeting, 
Smolenice, Czechoslovakia 1973}, eds. D. Krupa, J. Pisut (VEDA-Slovak Academy 
of Sciences, Bratislava, 1975) pp. 145-206.  
\item D. Zieminska, R.H. Dalitz, Decay processes 
$_{\Lambda}^8{\rm Li} \to \pi^- ~ ^8{\rm Be}$(17 Mev), Nucl. Phys. B 74 
(1974) 248-268. 
\item D. Bailin, R.V. Budny, G. Conforto, W.N. Cottingham, R.H. Dalitz, 
N. Dombey, D.H. Gabathuler, A. McDonald, D. Morgan, A.M. Segar, W.T. Toner, 
Physics with an electron-positron storage ring of beam energy $E=14$ GeV 
(RL-74-098, Chilton, 1974) 21 pp.
\item R.H. Dalitz, Heavy particles, in {\it Rudolf Peierls and Theoretical 
Physics, Proc. Oxford 1974}, eds. I.J.R. Aitchison, J.E. Paton (Pergamon, 
Oxford, 1977) pp. 86-100. 
\item R.H. Dalitz, Comments on heavy mesons, in {\it Experimental Meson 
Spectroscopy - 1974 (Boston)}, ed. D.A. Garelick, AIP Conf. Proc. 21 
(AIP, New York, 1974) pp. 75-78. 
\item D. Zieminska, R.H. Dalitz, Hypernucleus $_{\Lambda}^{12}{\rm B}$ and 
decay process $_{\Lambda}^{12}{\rm B} \to \pi^- + 3\alpha$, Nucl. Phys. A 238 
(1975) 453-490. 
\item R.H. Dalitz, Baryonic spectroscopy and its immediate future, in 
{\it New Directions in Hadron Spectroscopy, Proc. ANL Summer Symposium 1975}, 
eds. S.L. Kramer, E.L. Berger (Report ANL-HEP-CP-75-58, Argonne, 1975) 
pp. 383-406.
\item R.H. Dalitz, Quark physics, in {\it Few Body Dynamics, Proc. 7th Int'l 
Conf. Few Body Problems in Nuclear and Particle Physics, Delhi 1975}, eds. 
A.N. Mitra, I. Slaus, V.S. Bhasin, V.K. Gupta (North Holland, Amsterdam, 1976) 
pp. 632-658. 
\item R.H. Dalitz, The spectrum of baryonic states, in {\it Quarks and 
Hadronic Structure, Proc. Erice 1975}, ed. G. Morpurgo (Academic Press, 
New York, 1977) pp. 39-71. 
\item R.H. Dalitz, The status of noncharmed hadron spectroscopy, in {\it New 
Phenomena in Subnuclear Physics. Part B, Proc. Erice 1975}, ed. A. Zichichi 
(Academic Press, New York, 1977) pp. 1053-1066. 
\item R.H. Dalitz, A. Gal, Supersymmetric and strangeness analog states in
p-shell $\Lambda$ hypernuclei, Phys. Rev. Lett. 36 (1976) 362-365; Erratum, 
Phys. Rev. Lett. 36 (1976) 628. 
\item R.H. Dalitz, A. Gal, Quasi-free interactions in ($K^-,\pi^-$) 
strangeness-exchange nuclear-reactions at $0^o$, Phys. Lett. B 64 (1976) 
154-158; Erratum, Phys. Lett. B 97 (1980) 473. 
\item R.H. Dalitz, Quark models, old and new, in {\it Fundamentals of Quark 
Models, Proc. XVII Scottish Universities Summer School in Physics, St. Andrews 
1976}, eds. I.M. Barbour, A.T. Davies (Scottish Universities Summer School in 
Physics, Edinburgh, 1977) pp. 151-244. 
\item R.H. Dalitz, Introductory remarks on baryon spectroscopy, 
in {\it Baryon Resonances, Proc. Topical Conf. on Baryon Resonances, Oxford 
1976}, eds. R.T. Ross, D.H. Saxon (RHEL, Chilton, 1976) pp. 2-18. 
\item R.H. Dalitz, The conventional quark picture, in {\it Proc. Int'l 
Neutrino Conference, Aachen 1976}, eds. H. Faissner, H. Reithler, P.M. Zerwas, 
(Tech. Hochsch. Aachen, Braunschweig, 1977) pp. 451-463. 
\item R.H. Dalitz, Introductory remarks about new particles, Proc. Roy. Soc. 
A 355 (1977) 443-445. 
\item R.H. Dalitz, Charm and the $\psi$-meson family, Proc. Roy. Soc. A 355 
(1977) 601-619. 
\item R.H. Dalitz, Glossary for new particles and new quantum numbers, 
Proc. Roy. Soc. A 355 (1977) 629-631. 
\item R.H. Dalitz, R.R. Horgan, L.J. Reinders, The new resonance 
$\Delta$D35(1925) and ($56,1^-_3$) baryonic supermultiplet, J. Phys. G 3 
(1977) L195-L201. 
\item M. Jones, R.H. Dalitz, R.R. Horgan, Reanalysis of the baryon mass 
spectrum using the quark shell model, Nucl. Phys. B 129 (1977) 45-65. 
\item R.H. Dalitz, Quarks, colour and hadron spectroscopy, in {\it 
Topics in Quantum Field Theory and Gauge Theories, Proc. Salamanca 1977}, 
ed. J.A. de Azcarraga, Lecture Notes in Physics 77 (Springer-Verlag, Berlin, 
1978) pp. 336-378. 
\item R.H. Dalitz, L.J. Reinders, High lying baryonic multiplets in the 
harmonic quark shell model, in {\it Hadron Structure as Known from 
Electromagnetic and Strong Interactions, Proc. Hadron Structure '77 Conf., 
Strbske Pleso, Czechoslovakia}, ed. S. Dubnicka (VEDA, Bratislava, 1979) 
pp. 11-65. 
\item R.H. Dalitz, Hyperon-nucleon interactions and excited hypernuclei, in 
{\it Kaon-Nuclear Interaction and Hypernuclei, Proc. Seminar Zvenigorod 1977}, 
eds. P.A. Cherenkov et al. (Nauka, Moscow, 1979) pp. 355-384. 
\item R.H. Dalitz, A. Gal, Production and decay of $_{\Lambda}^7{\rm Li}$ 
and $_{\Lambda}^7{\rm He}$ hypernuclei, J. Phys. G 4 (1978) 889-906. 
\item A. Gal, J.M. Soper, R.H. Dalitz, Shell-model analysis of $\Lambda$ 
binding energies for p-shell hypernuclei. III. Further analysis and 
predictions, Ann. Phys. 113 (1978) 79-97. 
\item R.H. Dalitz, A. Gal, The formation of, and the $\gamma$-radiation from, 
the p-shell hypernuclei, Ann. Phys. 116 (1978) 167-243. 
\item R.H. Dalitz, The physics carried out with Nimrod, in {\it Nimrod The 
7 GeV proton synchrotron, Proc. Nimrod Commemoration Evening, Rutherford 
Laboratory 1978}, ed. J. Litt (SRC and Rutherford Lab., Chilton, 1979) 
pp. 40-53. 
\item R.H. Dalitz, Overview of $K^-$-nuclear reactions, in {\it Meson-Nuclear 
Physics - 1979 (Houston)}, ed. E.V. Hungerford III, AIP Conf. Proc. 54 
(AIP, New York, 1979) pp. 621-633. 
\item C.B. Dover, A. Gal, G.E. Walker, R.H. Dalitz, Angular distributions
for the $^{12}{\rm C}(K^-,\pi^-)_{\Lambda}^{12}{\rm C}$ reaction, 
Phys. Lett. B 89 (1979) 26-30. 
\item D. Kielczewska, D. Zieminska, R.H. Dalitz, Determination of the spin
of the hypernucleus $_{\Lambda}^{12}{\rm B}$. II. Complete analysis, 
Nucl. Phys. A 333 (1980) 367-380. 
\item R.H. Dalitz, C.R. Hemming, E.J. Morris, The $(K^-,\pi^-)$ strangeness 
exchange reactions on deuterium, Nukleonika 25 (1980) 1555-1580. 
\item R.H. Dalitz, Discrete $\Sigma$-hypernuclear states, Nature 285 (1980) 
11-12. 
\item R.H. Dalitz, J.G. McGinley, Remarks bearing on the interpretation of 
the $\Lambda(1405)$ resonance, in {\it Low and Intermediate Energy 
Kaon-Nucleon Physics}, eds. E. Ferrari, G. Violini (Reidel, Dordrecht, 1981) 
pp. 381-409. 
\item F.E. Close, R.H. Dalitz, The antisymmetric spin-orbit interaction 
between quarks, in {\it Low and Intermediate Energy Kaon-Nucleon Physics}, 
eds. E. Ferrari, G. Violini (Reidel, Dordrecht, 1981) pp. 411-418. 
\item R.H. Dalitz, Quarks in the context of few body physics, Nucl. Phys. 
A 353 (1981) 215c-232c. 
\item R.H. Dalitz, $\Lambda$-hypernuclear and $\Sigma$-hypernuclear physics, 
Nucl. Phys. A 354 (1981) 101c-125c. 
\item R.H. Dalitz, A. Gal, Strangeness analog states in 
$_{\Lambda}^9{\rm Be}$, Ann. Phys. 131 (1981) 314-355. 
\item R.H. Dalitz, Quarks and the light hadrons, in {\it Quarks and the 
Nucleus, Proc. Int'l School of Nucl. Phys. Erice 1981}, ed. D. Wilkinson, 
Prog. Part. Nucl. Phys. 8 (Pergamon, Oxford, 1982) pp. 7-48. 
\item R.H. Dalitz, Two relativistic effects for quarks within hadrons, 
Czech J. Phys. B 32 (1982) 597-608.  
\item R.H. Dalitz, A. Deloff, The $K^- {\rm D} \to p \Lambda \pi^-$ reaction, 
Czech J. Phys. B 32 (1982) 1021-1039. 
\item R.H. Dalitz, CP-nonconservation, Acta Phys. Austriaca Suppl. XXIV 
(1982) 393-474; Addendum, Acta Phys. Austriaca, 56 (1984) 130. 
\item R.H. Dalitz, Strange particle theory in the cosmic-ray period, 
in {\it Proc. Colloque Int'l sur l'Histoire de la Physique des Particules, 
Paris 1982}, J. Physique 43, Colloque C8 Suppl. (1982) 195-205; 406. 
\item R.H. Dalitz, J. McGinley, C. Belyea, S. Anthony, Theory of low-energy 
kaon-nucleon scattering, in {\it Proc. Int'l Conf. Hypernuclear and Kaon 
Physics, Heidelberg, 1982}, ed. B. Povh (MPI, Heidelberg, 1982) pp. 201-214.  
\item R.H. Dalitz, Hadron spectroscopy, in {\it Hawaii Topical Conference in 
Particle Physics, 1982}, eds. S. Pakvasa, S.F. Tuan (World Scientific, 
Singapore, 1982) Vol. 1, pp. 45-159. 
\item R.H. Dalitz, A. Deloff, The Strangeness-exchange reaction 
$K^- d \to p \Lambda \pi^-$ in flight, Australian J. Phys. 36 (1983) 
617-631. 
\item R.H. Dalitz, Recent perspectives in hadron spectroscopy, in {\it Proc. 
Hadron Structure '83, Smolenice}, ed. I. Lukac (Inst. Phys. EPRC, 
Bratislava, 1985) Vol. 1, pp. 17-46. 
\item R.H. Dalitz, Closing address, in {\it Proc. Hadron Structure '83, 
Smolenice}, ed. I. Lukac (Inst. Phys. EPRC, Bratislava, 1985) Vol. 2, 
pp. 533-539. 
\item R.H. Dalitz, C.H. Llewellyn Smith, Paying for high-energy physics, 
Nature 307 (1984) 104. 
\item R.H. Dalitz, Resonance - its description, criteria and significance, 
in {\it Resonances - Models and Phenomena, Proceedings Bielefeld 1984}, 
eds. S. Albeverico, L.S. Ferreira, L. Streit, Lecture Notes in Physics 211 
(Springer-Verlag, Berlin, 1984) pp. 1-26. 
\item R.H. Dalitz, Conference summary talk, in {\it The Intersections Between 
Particle and Nuclear Physics, Steamboat Springs 1984}, ed. R.E. Mischke, 
AIP Conf. Proc. 123 (AIP, New York, 1984) pp. 381-414. 
\item R.H. Dalitz, The $\tau - \theta$ puzzle, in {\it 50 Years of Weak 
Interactions: from the Fermi Theory to the W. Perspectives for the Future, 
Proc. Conf. Racine 1984}, eds. D. Cline, G.M. Reidasch (Univ. Wisconsin, 
Madison, 1984) pp. 332-348. 
\item D.J. Millener, A. Gal, C.B. Dover, R.H. Dalitz, Spin dependence of
the $\Lambda N$ effective interaction, Phys. Rev. C 31 (1985) 499-509. 
\item R.H. Dalitz, Reflections on Hideki Yukawa and the meson 50 symposium, 
Prog. Theor. Phys. Suppl. 85 (1985) 293-296. 
\item R.H. Dalitz, $K$-meson decays and parity violation, in {\it Pions to 
Quarks: Particle Physics in the 1950s, Proc. Second Int'l Symposium on the 
History of Particle Physics, Fermilab 1985}, eds. L.M. Brown, M. Dresden, 
L. Hoddeson (Cambridge University Press, Cambridge, 1989) pp. 434-457. 
\item R.H. Dalitz, S.F. Tuan, The continuing saga of the p wave singlet 
$\bar q q$ mesonic states, in {\it Proc. X Hawaii Conf. High Energy Physics, 
Honolulu 1985}, eds. F.A. Harris, S. Pakvasa, S.F. Tuan (University of Hawaii 
Press, Honolulu, 1986)
pp. 707-745.  
\item R.H. Dalitz, R.E. Peierls, Paul Adrien Maurice Dirac, Biogr. Mem. Fell. 
R. Soc. Lond. 32 (1986) 137-185; also in {\it Paul Adrien Maurice Dirac, 
1902-1984, elected F.R.S. 1930} (Cambridge University Press, Cambridge, 1986) 
pp. 139-185. 
\item R.H. Dalitz, D.H. Davis, D.N. Tovee, Proton decay of excited 
hypernuclei, Nucl. Phys. A 450 (1986) 311c-327c. 
\item R.H. Dalitz, P.I.P Kalmus, Quarks and leptons: the new elementary 
particles? preface, Proc. Roy. Soc. A 404 (1986) 151-152. 
\item M. Torres, R.H. Dalitz, A. Deloff, $K^-$ absorption reactions from 
rest in deuterium, Phys. Lett. B 174 (1986) 213-218. 
\item M. Torres, R.H. Dalitz, A. Deloff, The reactions 
$K^-d {\to} N \Lambda \pi$ and $N \Sigma \pi$, in {\it Intersections Between 
Particle and Nuclear Physics, Lake Louise 1986}, ed. D.F. Geesaman, AIP Conf. 
Proc. 150 (AIP, New York, 1986) pp. 901-905. 
\item M. Torres, R.H. Dalitz, A. Deloff, The strangeness-exchange reaction 
in deuterium, in {\it Proc. IX Oaxtepec Symposium on Nuclear Physics, 
Mexico 1986}, Notas de F\'{\i}sica 9 (1986) 367-386 . 
\item J. Hoek, R.H. Dalitz, Sweep-sweep correlations of Polyakov loops for
SU(3) lattice gauge theory on $32^4$ lattices, Phys. Lett. B 177 (1986) 
180-182. 
\item N.W. Tanner, R.H. Dalitz, The determination of T-violations and 
CPT-violations for the ($K^0,{\bar K}^0$) complex by ${\bar K}^0/K^0$ 
comparisons, Ann. Phys. 171 (1986) 463-488. 
\item R.H. Dalitz, Few body problems in strong interaction physics, 
Nucl. Phys. A 463 (1987) 37c. 
\item R.H. Dalitz, A biographical sketch of the life and work of Professor Sir 
Rudolf Peierls, FRS, in {\it A Breadth of Physics, Proc. Peierls 80th Birthday 
Symposium, Oxford 1987}, eds. R.H. Dalitz, R.B. Stinchcombe 
(World Scientific, Singapore, 1988) pp. 1-42. 
\item R.H. Dalitz, Professor Tony Hilton Royle Skyrme (1922-1987): A Brief 
Biography, in {\it A Breadth of Physics, Proc. Peierls 80th Birthday 
Symposium, Oxford 1987}, eds. R.H. Dalitz, R.B. Stinchcombe (World Scientific, 
Singapore, 1988) pp. 205-219. 
\item R.H. Dalitz, Another side to Paul Dirac, in {\it Paul Adrian Maurice 
Dirac: Reminiscences about a Great Physicist}, eds. B.N. Kursunoglu, 
E.P. Wigner (Cambridge University Press, Cambridge, 1987) pp. 69-92. 
\item R.H. Dalitz, A biographical sketch of the life of Professor 
P.A.M. Dirac, OM, FRS, in {\it Tributes to Paul Dirac}, ed. J.G. Taylor 
(IOP, Bristol, 1987) pp. 3-28. 
\item R.H. Dalitz, Historical remark on $\pi^0 \to \gamma e^+ e^-$ decay, 
in {\it 40 Years of Particle Physics, Proc. Int'l Conf. to Celebrate the 40th 
Anniversary of the Discovery of the $\pi$- and V-particles, Bristol 1987}, 
eds. B. Foster, P.H. Fowler (Adam Hilger, Bristol, 1988) pp. 105-108. 
\item R.H. Dalitz, The pion in particle physics today - an overview, 
in {\it 40 Years of Particle Physics, Proc. Int'l Conf. to Celebrate the 40th 
Anniversary of the Discovery of the $\pi$- and V-particles, Bristol 1987}, 
eds. B. Foster, P.H. Fowler (Adam Hilger, Bristol, 1988) pp. 157-188. 
\item R.H. Dalitz, Proton proton and anti-proton proton scattering at the 
highest energies, in {\it New Aspects of High-Energy Proton-Proton Collisions, 
Proc. 4th INFN ELOISATRON Project Workshop, Erice 1987}, ed. A. Ali 
(Plenum, New York, 1988) pp. 281-308. 
\item I.J. Ford, R.H. Dalitz, J. Hoek, Potentials in pure QCD on $32^4$ 
lattices, Phys. Lett. B 208 (1988) 286-290. 
\item S.F. Tuan, T. Ferbel, R.H. Dalitz, Comments on the evidence for 
a $1^{-+}$ exotic meson, Phys. Lett. B 213 (1988) 537-540. 
\item R.H. Dalitz, An outline of the life and work of Tony Hilton Royle Skyrme 
(1922-1987), Int. J. Mod. Phys. A 3 (1988) 2719-2744; also in {\it Selected 
Papers, with Commentary, of Tony Hilton Royle Skyrme}, ed. Gerald E. Brown, 
World Scientific Series in 20th Century Physics - Vol. 3 (World Scientific, 
Singapore, 1994) pp. 6-31. 
\item R.H. Dalitz, G.R. Goldstein, R. Marshall, Heavy quark spin correlations 
in $e^+e^-$ annihilations, Phys. Lett. B 215 (1988) 783-787. 
\item R.H. Dalitz, Quantum chromodynamics and excited baryons, 
in {\it Excited Baryons 1988, Proc. Troy 1988}, eds. G. Adams, 
N.C. Mukhopadhyay, P. Stoler (World Scientific, Singapore, 1989) pp. 525-558. 
\item G.L. Shaw, M. Shin, R.H. Dalitz, M. Desai, Growing drops of strange 
matter, Nature 337 (1989) 436-439. 
\item R.H. Dalitz, G.R. Goldstein, R. Marshall, On the helicity of charm 
jets, Z. Phys. C 42 (1989) 441-448. 
\item R.H. Dalitz, D.H. Davis, P.H. Fowler, A. Montwill, J. Pniewski, 
J.A. Zakrzewski, The identified $\Lambda\Lambda$-hypernuclei and the predicted 
$H$-particle, Proc. Roy. Soc. A 426 (1989) 1-17. 
\item R.H. Dalitz, D.H. Davis, A. Deloff, Is there a bound $_{\Sigma}^4$He?, 
Phys. Lett. B 236 (1990) 76-80. 
\item R.H. Dalitz, Neutral kaon beams, in {\it Plots, Quarks and Strange 
Particles, Proc. Dalitz Conf. 1990}, eds. I.J.R.~Aitchison, C.H. Llewellyn 
Smith, J.E. Paton (World Scientific, Singapore, 1991) pp. 118-133. 
\item R.H. Dalitz, A. Deloff, The shape and parameters of the $\Lambda$(1405) 
resonance, J. Phys. G 17 (1991) 289-302; Erratum, J. Phys. G 19 (1993) 1423.  
\item R. H. Dalitz, Andrei Sakharov, scientific dissertation, 
Usp. Fiz. Nauk 161 (1991) 121-136. 
\item R. H. Dalitz, The Young Sakharov's ``Isotopic Parity", 
in {\it Sakharov Memorial Lectures in Physics, Proc. 1st Sakharov 
Conf. on Physics, 21-31 May 1991}, ed. L.V.~Keldysh, V.Ya.~Fainberg 
(Nova Science Publ., Commack N.Y., 1992) pp. 635-666. 
\item R. H. Dalitz, The discrete symmetries (C,P,T) and their exploration 
with neutral kaons, Nucl. Phys. B Proc. Suppl. 24 (1991) 3-23. 
\item R.H. Dalitz, Some themes from physics in collision, in {\it Physics in 
Collision 11, Proc. Colmar 1991}, eds. J.-M. Brom, D. Huss, M.-E. Michalon 
(Editions Frontieres, Paris, 1991) pp. 535-561.  
\item R.H. Dalitz, G.R. Goldstein, Decay and polarization properties of the 
top quark, Phys. Rev. D 45 (1992) 1531-1543. 
\item R.H. Dalitz, G.R. Goldstein, Analysis of top-antitop production and 
dilepton decay events and the top quark mass, Phys. Lett. B 287 (1992) 
225-230. 
\item R.H. Dalitz, A. Deloff, Is bound $_{\Sigma}^4$He formed in $K^-$-meson 
capture from rest in liquid-helium?, Nucl. Phys. A 547 (1992) 181c-190c. 
\item R.H. Dalitz, Top quark, New Scientist 135 (1992) 47. 
\item R.H. Dalitz, Poles, cusps and other singularities, in {\it Strangeness 
in Nuclei, Proc. Workshop Cracow 1992}, eds. S. Kistryn, O.W.B. Schult 
(World Scientific, Singapore, 1993) pp. 203-249. 
\item R.H. Dalitz, G.R. Goldstein, Where is Top?, in {\it From Superstrings 
to the Real World, Proc. Int'l School of Subnuclear Physics, Erice 1992}, 
ed. A. Zichichi (World Scientific, Singapore, 1993) pp. 442-467. 
\item G.R. Goldstein, K. Sliwa, R.H. Dalitz, A technique for observing the top 
quark and measuring its mass at the Tevatron, in {\it Proc. XXVI Int'l Conf. 
High Energy Physics, Dallas 1992}, ed. J.R. Sanford, AIP Conf. Proc. 272 
(AIP, New York, 1993) pp. 1027-1030. 
\item G.R. Goldstein, K. Sliwa, R.H. Dalitz, Observing top-quark production 
at the Fermilab Tevatron, Phys. Rev. D 47 (1993) 967-972. 
\item R.H. Dalitz, The $\tau - \theta$ puzzle, in {\it Discovery of Weak 
Neutral Currents: The Weak Interaction Before and After, Santa Monica 1993}, 
eds. A.K. Mann, D.B. Cline, AIP Conf. Proc. 300 (AIP, New York, 1993) 
pp. 141-158. 
\item R.H. Dalitz, G.R. Goldstein, Where is Top? (based on a talk given by 
R.H. Dalitz at Erice 1992) Int. J. Mod. Phys. A 9 (1994) 635-666. 
\item R.H. Dalitz, A. Deloff, Is bound $_{\Sigma}^4$He formed in $K^-$-meson 
capture from rest in liquid-helium?: II, Nucl. Phys. A 585 (1995) 303c-306c. 
\item R.H. Dalitz, Kaon decays to pions, The $\tau - \theta$ problem, in 
{\it History of Original Ideas and Basic Discoveries in Particle Physics, 
Proc. NATO Adv. Workshop, Erice 1994}, eds. H.B. Newman, T. Ypsilantis 
(Plenum Press, New York, 1996) pp. 163-181; 183. 
\item R.H. Dalitz, Complete bibliography for Prof. Sir Rudolf Peierls, 
Nucl. Phys. A 604 (1996) 7-23. Erratum, Nucl. Phys. A 608 (1996) 513. 
\item R.H. Dalitz, Some personal recollections of S. Chandrasekhar at Chicago 
and Oxford, in {\it S. Chandrasekhar: the Man Behind the Legend, Chandra 
Remembered}, ed. K.C. Wali (Imperial College Press, London, 1997) pp. 142-155.  
\item R.H. Dalitz, D.H. Davis, T. Motoba, D.N. Tovee, Proton emitting 
$\Lambda$-bound states of $_{\Lambda}^{16}{\rm O}^{\star}$, Nucl. Phys. A 625 
(1997) 71-94. 
\item R.H. Dalitz, D.H. Davis, T. Motoba, D.N. Tovee, Proton emitting 
$\Lambda$-bound states of hypernuclei, in {\it Proc. Eur. Conf. Advances in 
Nuclear Physics and Related Areas, Thessaloniki 1997}, eds. D.M. Brink, 
M.E. Grypeos, S.E. Massen (Giahoudi-Giapouli, Thessaloniki, 1999) pp. 582-592.  
\item R.H. Dalitz, Note on the $\Lambda$(1405), in {\it Review of Particle 
Physics by the Particle Data Group}, eds. C. Caso et al., Eur. Phys. J. C 3 
(1998) 676-678. 
\item R.H. Dalitz, G.R. Goldstein, Test of analysis method for top-antitop 
production and decay events, Proc. Roy. Soc. A 455 (1999) 2803-2834. 
\item R.H. Dalitz, Kaon physics: the first 50+ years, in {\it Kaon Physics, 
Proc. KAON'99}, eds. J.L. Rosner, B.D. Winstein (University of Chicago Press, 
Chicago, 2001) pp. 5-22. 
\item R.H. Dalitz, G. Garbarino, Local realistic theories and quantum 
mechanics for the two-neutral-kaon system, Nucl. Phys. B 606 (2001) 483-517.
\item R.H. Dalitz, Hypernuclear physics as we enter the third millenium, 
Nucl. Phys. A 691 (2001) 1c-10c. 
\item R. Dalitz, Paul Dirac: a genius in the history of physics, CERN Courier 
42(7) (2002) 15-17. 
\item R.H. Dalitz, 50 years of hypernuclear physics. II. The later years, 
Nucl. Phys. A 754 (2005) 14c-24c. 

\end{enumerate} 

\vspace*{1cm} 

\section{Authored book reviews, public lectures and obituaries, 
and books edited by R.H. Dalitz}
\noindent 
{\it by I.J.R. Aitchison, F.E. Close, A. Gal and D.J. Millener} 
 
\begin{enumerate}

\item R.H. Dalitz, book review on The Eightfold Way, by M. Gell-Mann and 
Yuval Ne'eman, Endeavour 24 (1965) 168. 

\item R.H. Dalitz, book review on Nuclear Interactions, by Sergio DeBenedetti, 
J. Franklin Inst. - Eng. and Appl. Math. 280(3) (1965) 280. 

\item R.H. Dalitz, book review on Unitary Symmetries and their Applications 
to High Energy Physics, by M. Gourdin, J. Franklin Inst. - Eng. and Appl. 
Math. 284(6) (1967) 431. 

\item W.K.H. Panofsky, R.H. Dalitz, Particle Physics (thirteen chapters) 
Thirteen chapters from {\it Nuclear Energy Today and Tomorrow: a course of 
lectures on selected topics in the fields of nuclear and atomic energy}, 
delivered in the Twelfth Int'l Science School for High School Students 
sponsored by the Science Foundation for Physics within the University of 
Sydney, 1969, eds. H. Messel, S.T. Butler (Heinemann Education Books, London, 
1971) pp. 313-473.  

\item R.H. Dalitz, ed., A discussion on duality: reggeons and resonances in 
elementary particle processes, {\it Proc. Royal Society Discussion Meeting 
1970}, Proc. Roy. Soc. A 318 (1970) 243-399. 

\item R.H. Dalitz, book review on Spectroscopic and Group Theoretical Methods 
in Physics, edited by F. Bloch, S.G. Cohen, A. De-Shalit, S. Sambursky 
and I. Talmi, J. Franklin Inst. - Eng. and Appl. Math. 289(4) (1970) 323-324. 

\item R.H. Dalitz, book review on High Energy Physics, Vol. III, edited by 
E. H. S. Burshop, J. Franklin Inst. - Eng. and Appl. Math. 291(2) (1971) 
145-146. 

\item R.H. Dalitz, A. Zichichi, eds., Meson Resonances and Related 
Elecromagnetic Phenomena, Proc. 1st Int'l Conf. organized by the High Energy 
and Particle Physics Division of the EPS, Bologna 1971 (Editrice Compositori, 
Bologna, 1972) 624 pp. 

\item R.H. Dalitz, book review on Theoretical physics -- I and II, 
Vol. I, by Benjamin Levich, J. Franklin Inst. - Eng. and Appl. Math. 298(1) 
(1974) 75-77. 

\item R.H. Dalitz, book review on High Energy Physics, Vol. V, edited by 
E.H.S. Burhop, J. Franklin Inst. - Eng. and Appl. Math. 299(2) (1975) 147-148. 

\item R.H. Dalitz, B. Richter, B. Wiik, W.T. Toner, eds., A discussion on new 
particles and new quantum numbers, {\it Proc. Royal Society Discussion Meeting 
1977}, Proc. Roy. Soc. A 355 (1977) 441-631.

\item R.H. Dalitz, Style of a physicist: Tabibito (The Traveler) by Hideki 
Yukawa, Nature 302 (1983) 770-771. 

\item R.H. Dalitz, Preface to {\it Hadron Interactions} by P.D.B. Collins, 
A.D. Martin (Adam Hilger, Bristol, 1984) 3 pp. 

\item R.H. Dalitz, All of physics: McGraw-Hill Encyclopedia of Physics by 
S.P. Parker, Nature 307 (1984) 393. 

\item R.H. Dalitz, Recollections of physics: The Birth of Particle Physics by 
L.M. Brown and L. Hoddeson, Nature 308 (1984) 383-384. 

\item R.H. Dalitz, Elusive particles, Times Higher Educational Supplement, 
issue of 15 March 1985, p.~29.

\item R.H. Dalitz, P.I.P. Kalmus, eds., Quarks and Leptons: the new elementary 
particles? {\it Proc. Royal Society Discussion Meeting 1985}, Proc. Roy. Soc. A 
404 (1986) 151-298. 

\item R.H. Dalitz, Fundamental developments: Constructing Quarks - 
a Sociological History of Particle Physics by A. Pickering, Nature 314 (1985) 
387-388. 

\item R.H. Dalitz, Beyond hope, Bohr and physics: Niels Bohr - a Centenary 
Volume by A.P. French, P.J. Kennedy, Nature 320 (1986) 221-222. 

\item R.H. Dalitz, R.B. Stinchcombe, eds., A Breadth of Physics, Proceedings 
of the Peierls 80th Birthday Symposium, Oxford University, 27 June 1987 
(World Scientific, Singapore, 1988) 234 pp. 

\item R.H. Dalitz, E.L. Hahn, Professor Emilio Segre, The Independent, 
issue of 8 April 1989, p.~1. 

\item R.H. Dalitz, Emilio Gino Segre 1905-89, Physics World 2(2) (1989) 
57-58.

\item R.H. Dalitz, Into the atom: The Atomic Scientists. A Biographical 
History by H.A. Boorse, L. Motz and J.H.~Weaver, Nature 344 (1990) 898. 

\item R.H. Dalitz, Dirac Paul Adrien Maurice, in {\it Dict. Nat. Bio. 1981-5}, 
eds. Lord Blake, C.S. Nicholls (Oxford University Press, Oxford, 1990) 
pp.~115-116. 

\item R.H. Dalitz, A.D. Sakharov (1921-1989), Physics World 3(2) (1990) 55-56. 

\item R.H. Dalitz, Dirac Paul Adrien Maurice, in {\it Bio. Dict. Scientists}, 
ed. Trevor I. Williams (HarperCollins, Glasgow, 1993) pp.~140-141.  

\item R.H. Dalitz, ed., The Collected Works of P.A.M. Dirac: 1924-1948, 
Vol. 1, (Cambridge University Press, Cambridge, 1995) 1310 pp. 

\item R.H. Dalitz, At the Heart of the Atom [Obituary of Sir Rudolf Peierls], 
The Guardian, issue of 21 September 1995, p.~15. 

\item R.H. Dalitz, A Hundredth Happy Birthday, The Times Higher Education 
Supplement, issue of 8 December 1996, p.~30.  

\item R.H. Dalitz, Sir Rudolf Peierls, eds., Selected Scientific Papers of Sir 
Rudolf Peierls With Commentary (World Scientific, Singapore, 1997) 805 pp.

\item R.H. Dalitz, Mott and the cosmic radiation, in {\it Nevill Mott: 
Reminiscences and Appreciations}, ed. E.A. Davis (Taylor \& Francis, London, 
1998) pp.~85-90. 

\item R.H. Dalitz, Nicholas Kemmer 1911-1998, Physics World 12(1) (1999) 43. 

\item R.H. Dalitz, N. Dombey, Peierls was no spy, Spectator, issue of 10 July 
1999, p. 23. 

\item R.H. Dalitz, Sir Rudolf Peierls was not a spy, Physics World 12(10) 
(1999) 15-16. 

\item R.H. Dalitz, M. Nauenberg, eds., The Foundations of Newtonian Scholarship 
(World Scientific, Singapore, 2000) 260 pp. 

\item R.H. Dalitz, F.J. Duarte, John Clive Ward - Obituary, Physics Today 
53(10) (2000) 99-100. 

\item R.H. Dalitz, Messages from Founding Fellows, Churchill Review, 
37 (2000) 17. 

\item R.H. Dalitz, Marcus Laurence Elwin `Mark' Oliphant - Obituary, Physics 
Today 54(7) (2001) 73-74. 

\item R.H. Dalitz, Philip Burton Moon, in {\it Oxford - DNB}, eds. Brian 
Harrison, H.C.G. Matthew, vol. 38 (Oxford University press, Oxford, 2004) 
pp.~896-897. 

\item R.H. Dalitz, Rudolf Ernst Peierls, in {\it Oxford - DNB}, eds. Brian 
Harrison, H.C.G. Matthew, vol. 43 (Oxford University press, Oxford, 2004) 
pp.~442-445. 

\end{enumerate}


\normalsize

\begin{thebibliography}{99} 

\bibitem{Dal51} R.H. Dalitz, Proc. Phys. Soc. A 64 (1951) 667. 

\bibitem{Pla59} R. Plano et al. Phys. Rev. Lett. 3 (1959) 525. 

\bibitem{Cou63} H. Courant et al. Phys. Rev. Lett. 10 (1963) 409. 

\bibitem{Dal53} R.H. Dalitz, Phil. Mag. 44 (1953) 1068.  

\bibitem{Dal54} R.H. Dalitz, Phys. Rev. 94 (1954) 1046. 

\bibitem{Dal57} R.H. Dalitz, Rep. Prog. Phys. 20 (1957) 163. 

\bibitem{Dal61} R.H. Dalitz, Rev. Mod. Phys. 33 (1961) 471. 

\bibitem{Dal63} R.H. Dalitz, Annu. Rev. Nucl. Sci. 13 (1963) 339. 

\bibitem{Dal59} R.H. Dalitz, S.F. Tuan, Phys. Rev. Lett. 2 (1959) 425. 

\bibitem{Als61} M. Alston et al., Phys. Rev. Lett. 6 (1961) 698. 

\bibitem{Als60} M. Alston et al., Phys. Rev. Lett. 5 (1960) 520. 

\bibitem{DWR67} R.H. Dalitz, T.C. Wong, G. Rajasekaran, Phys. Rev. 153 (1967) 
1617. 

\bibitem{CDD56} L. Castillejo, R.H. Dalitz, F.J. Dyson, Phys. Rev. 101 (1956) 
453. 

\bibitem{CDa81} F.E. Close, R.H. Dalitz, in {\it Low and Intermediate Energy 
$KN$ Physics}, eds. E. Ferrari, G. Violini (Reidel, Dordrecht, 1981) 411. 

\bibitem{BDa63} M.M. Block, R.H. Dalitz, Phys. Rev. Lett. 11 (1963) 96. 

\bibitem{DGa78} R.H. Dalitz, A. Gal, Ann. Phys. 116 (1978) 167. 

\bibitem{MGD85} D.J. Millener, A. Gal, C.B. Dover, R.H. Dalitz, 
Phys. Rev. C 31 (1985) 499. 

\bibitem{HTa06} O. Hashimoto, H. Tamura, Spectroscopy of $\Lambda$ 
hypernuclei, Prog. Part. Nucl. Phys., in press. 

\bibitem{DGa76} R.H. Dalitz, A. Gal, Phys. Rev. Lett. 36 (1976) 362. 

\bibitem{Dal05} R.H. Dalitz, Nucl. Phys. A 754 (2005) 14c. 

\bibitem{DGo92} R.H. Dalitz, G.R. Goldstein, Phys. Rev. D 45 (1992) 1531. 

\end{thebibliography}
\end{document}